\documentclass[sigconf]{acmart}
\AtBeginDocument{%
  }

\copyrightyear{2026}
\acmYear{2026}
\setcopyright{cc}
\setcctype{by-nc-nd}
\acmConference[CHI EA '26]{Extended Abstracts of the 2026 CHI Conference on Human Factors in Computing Systems}{April 13--17, 2026}{Barcelona, Spain}
\acmBooktitle{Extended Abstracts of the 2026 CHI Conference on Human Factors in Computing Systems (CHI EA '26), April 13--17, 2026, Barcelona, Spain}
\acmDOI{10.1145/3772363.3798839}
\acmISBN{979-8-4007-2281-3/2026/04}

\begin{document}

\title{Sticky and Magnetic: Evaluating Error Correction and User Adaptation in Gaze-and-Pinch Interaction}


\author{Jazmin Collins}
\affiliation{%
  \institution{Cornell University}
  \city{New York, NY}
  \country{USA}}
\email{jc2884@cornell.edu}

\author{Prasanthi Gurumurthy}
\affiliation{%
  \institution{Google}
  \city{San Jose}
  \country{USA}}
\email{gprasanthi@google.com}

\author{Eric J. Gonzalez}
\affiliation{%
 \institution{Google}
 \city{Seattle}
 \country{USA}}
\email{ejgonz@google.com}

\author{Mar Gonzalez-Franco}
\affiliation{%
  \institution{Google}
  \city{Seattle}
  \country{USA}}
\email{margon@google.com}

\renewcommand{\shortauthors}{Trovato et al.}

\begin{abstract}
  The gaze-and-pinch framework offers a high-fidelity interaction modality for spatial computing in virtual reality (VR), yet it remains vulnerable to coordination errors—timing misalignments between gaze fixation and pinch gestures. These errors are categorized into two types: late triggers (gaze leaves a target before pinch) and early triggers (pinch before gaze arrival on target). While late triggers are well-studied, early triggers lack robust solutions. We investigate two heuristics—\textsc{Sticky} selection (temporal buffer) and \textsc{Magnetic} selection (spatial field)—to mitigate these errors. A within-subjects study ($N=9$) on the Samsung Galaxy XR evaluated these heuristics against a baseline. Findings indicate that while throughput and selection time remained stable, the heuristics fundamentally shifted user behavior and significantly reduced errors during selection. Notably, \textsc{Magnetic} selection induced an ``offloading'' effect where users traded precision for speed. Additionally, the heuristics reclassified ambiguous failures as explainable coordination errors. We provide recommendations for selection heuristics that enhance interaction speed and cognitive agency in virtual reality.
\end{abstract}


\begin{teaserfigure}
\centering
 \includegraphics[width=\textwidth]{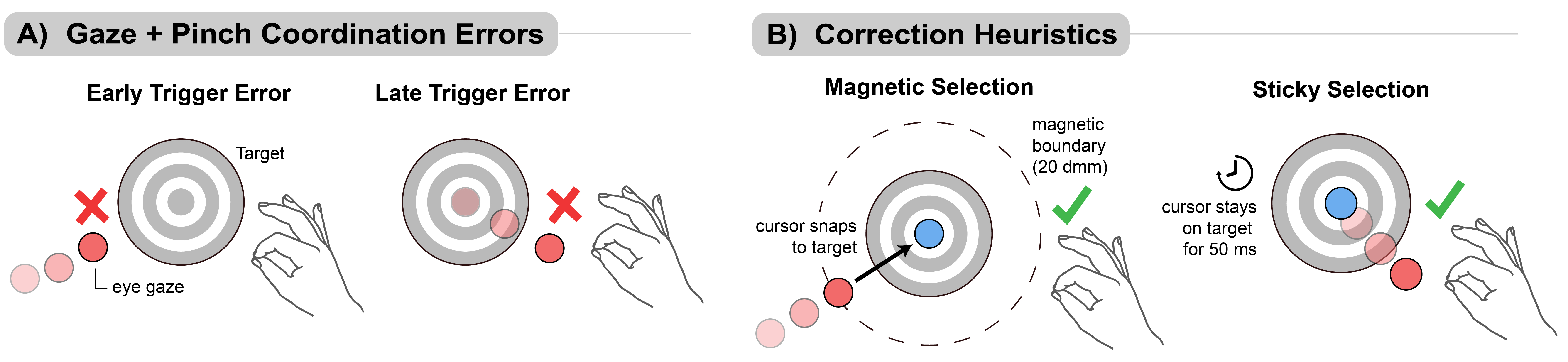}
  \caption{We investigate two types of coordination errors common to gaze-and-pinch interactions (A), Early and Late Trigger errors, along with two correction heuristics (B), Magnetic and Sticky selection.}
  \label{fig: heuristics}
  \Description{}
\end{teaserfigure}

\maketitle{}

\section{Introduction}
Virtual reality (VR) is growing in popularity across sectors, from healthcare to education. As major firms continue to invest in the ecosystem, VR is transitioning from entertainment technology into a platform for spatial computation. In this new paradigm, a 3D digital workspace promises immersion and utility that could supersede traditional computing.

However, realizing the potential of spatial computing requires input methods that match the fidelity of traditional interfaces. In particular, making selections in VR remains a key usability bottleneck. Users find that selecting objects in 3D space is slower, less accurate, and more fatiguing than using standard devices such as a mouse or trackpad \cite{kargut2024effects, barrett2022comparing, spiess2023comparison}. To address this, the HCI community has examined multimodal interactions in VR \cite{vanacken2009multimodal, kim2000multimodal, mutasim2021pinch, kim2025pinchcatcher}, including the gaze-and-pinch framework \cite{pfeuffer2017gaze+}, which combines the speed of eye tracking for targeting with the ease of pinch gestures for confirmation.

However, gaze-and-pinch interaction still poses challenges to users. In particular, prior work has noted that because gaze is a highly rapid input—often moving faster than manual motor responses—users frequently experience \textbf{coordination errors} when performing gaze-and-pinch selection \cite{pfeuffer2024design, park2024impact, park2025gazehandsync}. Such coordination errors (i.e., when the timing of the gaze fixation and the pinch do not align) are particularly common due to the asynchrony between high-speed gaze and the slower, manual pinch. These errors fall into two categories, shown in Figure \ref{fig: heuristics}A:
\textbf{Late Trigger Errors}, which occur when a user fixates on a target but shifts their gaze away before they pinch, and \textbf{Early Trigger Errors}, which occur when a user initiates the pinch before their gaze lands on the target.

Past work has attempted to address these errors, exploring solutions to late trigger errors like machine learning (ML) models trained to identify late pinches and correct system output \cite{park2025gazehandsync}. However, early trigger errors remain underexplored, despite major differences between the error types. With late trigger errors, a system can review past gaze data to identify and correct for an intended target; in contrast, early triggers require a predictive approach to determine where gaze was \textit{going} when the pinch occurred. Solutions for late and early triggers are fundamentally incompatible, leaving a gap in how we address them both. We aimed to address this gap by investigating two solutions for coordination errors (Figure \ref{fig: heuristics}B): 

\begin{itemize}
    \item (1) \textbf{\textsc{Sticky} selections}: A temporal adjustment that keeps a gaze reticle artificially ``stuck'' to a target for a brief window of time (designed for late trigger errors).
    \item (2) \textbf{\textsc{Magnetic} selections}: A spatial adjustment that adds a larger ``magnetic field'' around an intended target, snapping a gaze reticle to the center of the target once it touches the field (designed for early trigger errors).
\end{itemize}

We evaluated these two heuristics in a within-subjects study with nine expert users of gaze-and-pinch. We had our participants perform repeated target selections in VR with a baseline and each of our heuristics. Finally, we collected log data of their performances to evaluate the effects of our heuristics via statistical analysis of key metrics. Ultimately, we discovered that our heuristics significantly reduced the errors that participants would have experienced during selection, without impacting performance metrics such as throughput, accuracy, or selection time. We also identified behavioral shifts among our participants as they adapted to the heuristics, such as performing less precise selections on targets with a "magnetic field." By sharing these results, we provide recommendations for mitigation of coordination errors to create a truly seamless spatial computing experience in VR.

\section{Related Work}

\subsection{Gaze-and-Pinch Challenges}

The concept of combining gaze for coarse targeting with manual input for fine selection was first popularized by Zhai et al. with MAGIC pointing \cite{zhai1999magic}. Since then,HCI researchers have explored various interaction problems with gaze-and-pinch selections in VR \cite{park2025gazehandsync, pfeuffer2024design, park2024impact, pfeuffer2017gaze+}. Notably, Pfeuffer et al. \cite{pfeuffer2024design} presented five principles of gaze-and-pinch interactions and challenges that must be addressed, such as improving drag and drop mechanics, expanding how users manipulate objects, and addressing coordination errors. In recent years, the HCI community has begun to explore one of these challenges, coordination errors, in more depth \cite{park2025gazehandsync, park2024impact}. Park et al. trained ML models to identify late triggers, incorporating dynamic timing adjustments to pinches that were flagged as late \cite{park2025gazehandsync}. These models resulted in an overall decrease in late triggers among their participant group, demonstrating a promising approach for these errors.

While previous work has explored solutions for \textit{late} trigger errors, there remains a gap in the study of \textit{early} trigger errors. Although less frequent, early triggers pose a distinct usability barrier in gaze-and-pinch interaction. Specifically, late trigger errors are addressed through retroactive adjustment, leveraging past gaze data to identify the intended target. In contrast, early triggers occur before the user’s gaze has reached the target, necessitating predictive action. Thus, there is a need to explore unique solutions for early trigger errors. Our work has taken a first step to this goal, introducing two heuristics designed to compensate for early and late triggers, respectively.

\subsection{Cognitive Impact of Coordination Errors}

Within the neuroscience community, the concept of intentional binding, or action-binding, provides a critical framework for understanding coordination errors. Action-binding describes the process of mentally linking an action that one intended to perform to its result \cite{haggard2003intentional, haggard2003conscious, haggard2002voluntary, chambon2015tms, obhi2011sense, schwarz2018action}. Past work in this area by Haggard et al. has shown that when results following intentional actions deviate from expectations, the user experiences diminished agency \cite{haggard2003intentional}. Earlier work by Chambon et al. supported this, determining that a person's sense of agency over their actions can be significantly reduced by the stimulation of involuntary movements interrupting their actions \cite{chambon2015tms}.

This erosion of agency underscores why coordination errors are more than technical glitches; they are fundamental breaks in the user’s cognitive loop. For a VR system to feel intuitive, the user must experience consistent action-binding, where intentional gestures yield expected results. If there are too many disruptions within the VR system, this state of action-binding will not be reached and users may feel as though they or the system is not working as expected, causing frustration. Late or early trigger errors are prime examples of such disruptions in VR, resulting in incorrect selections, unintentional interactions with irrelevant objects, or delays in navigating through VR interfaces as users repeatedly ``miss'' intended targets. By proposing solutions to mitigate these errors, we introduce a method to preserve action-binding and improve the overall usability of gaze-and-pinch selection in VR.

\section{Methods}

\subsection{Participants}

Participants were recruited from an internal roster at Google that performs weekly VR tasks involving gaze-and-pinch selection and text entry. Fifteen participants were initially recruited. However, only nine participants' data was retained due to technical failures (e.g., logging errors and file corruption during remote sending of data). While this resulted in an uneven distribution across our counterbalancing, crucially, these exclusions were purely technical and independent of participant performance. Consequently, we treated the study as a preliminary investigation and focused our analysis on initial differences across conditions and users' behavioral shifts. All procedures were approved by Google's internal ethics board.

\subsection{Prototype Design and Procedure}

The study utilized the Samsung Galaxy XR headset, with a custom evaluation prototype developed in Unity. We designed two heuristics to address distinct coordination errors (Figure \ref{fig: heuristics}):

\begin{itemize}
    \item \textbf{\textsc{Sticky} Selection}: To mitigate late triggers, the gaze reticle remains artificially fixed to a target for $50~ms$ after the user's actual gaze departs.
    \item \textbf{\textsc{Magnetic} Selection}: To mitigate early triggers, a $20~dmm$ (distance-independent millimeter, a unit of measurement in VR that gives 2D UI elements a consistent visual size \cite{googledmm}) ``magnetic field'' was added to each target. If a pinch occurs while the gaze is within this field, the selection snaps to the target center.
\end{itemize}

\begin{figure}
\includegraphics[width=\columnwidth]{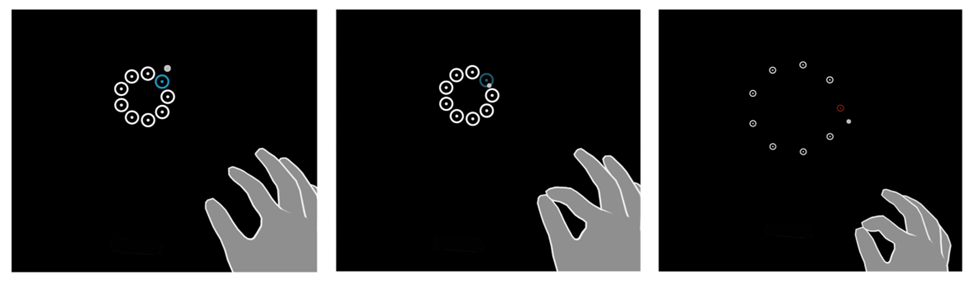}
\caption{\label{fig: targetsetup} Target setup in the selection task showing correct (middle) and incorrect (right) selections.}
\Description{ Three images depict the virtual environment that the selection task took place in. Each image shows nine circular targets laid out in a larger circular pattern, equidistant from each other and the center of the circle. The first two images show circles of the same size and layout, while the third image shows circles of a much smaller size spread farther apart. A gray outline of a user's hand is visible in the lower right corner, showing whether their hand is open or pinching to select.

In the first two images, a white circle representing the user's gaze is nearing a target highlighted in blue. In the second image, the white circle is laid over the blue target and the user's hand is in a pinch gesture. The white circle has shrunk and the blue target has become a darker shade of blue to indicate it is being selected. In the third image, the user's hand is pinching to select, but the white circle of their gaze is off the highlighted target. The target itself is also colored a dark shade of red, indicating that the user missed the target and made an incorrect selection.
}
\vspace{-15pt}
\end{figure}

We evaluated these heuristics in a 2D target selection task in VR. The virtual environment contains a ring of nine equally spaced circular targets of the same size (see Figure \ref{fig: targetsetup}) based on the standard 2D Fitts' Law task \cite{fittsiso20129241}. The targets are located in a vertical plane 1.3 m from the user. Inter-target distance (between adjacent targets) varies between target setups and is either 0.13 m ($\approx5.72^\circ$ visual angle) or 0.26 m ($\approx11.42^\circ$ visual angle). In each trial, a target is highlighted and the user attempts to select it by looking at it and pinching. The target flashes to indicate if they succeeded or failed, and the next target is highlighted. After nine selections have been made, a new round of targets of different sizes (alternating between visual angles of 1.43°, 2.03°, 2.86°, 4.05°, and 5.72°) is presented. When all selections are complete, the environment closes.

Our study procedure was facilitated remotely by a study coordinator. Participants were given an APK, an instructional document, and a bash script to toggle study conditions. After performing the Galaxy XR’s built-in eye-tracking calibration, participants completed four blocks of selection tasks. Each block consisted of 10 rounds of 9 targets, totaling 90 selections per block. The four blocks corresponded to our conditions: (1) \textsc{None} (Baseline, no heuristics added), (2) \textsc{Sticky}, (3) \textsc{Magnetic}, and (4) \textsc{Sticky+Magnetic} (where both heuristics were applied). These were counterbalanced to prevent order effects. At the conclusion of the session, the prototype generated frame-by-frame and selection-specific logs for analysis.

\subsection{Data and Analysis}

We conducted a two-stage analysis of the study data. In total, we analyzed 72 data logs (eight logs per participant).

For the first stage of our analysis, we processed the raw logs to calculate six variables: Throughput (task difficulty / movement time of a user's gaze to hit a target), Error Rate (percentage of incorrect selections), Late/Early Trigger Error Rates (percentage of error rates attributed to late or early trigger errors), Average Selection Time (average time in milliseconds to make a selection), and Error Reduction. To calculate Error Reduction, we tracked ``would-be'' errors -- selections that would have failed without a heuristic (based on raw gaze position and pinch data) but were corrected by the \textsc{Sticky} or \textsc{Magnetic} logic. Late and early trigger errors were classified using a $350~ms$ temporal window (pinches occurring within $350~ms$ before a gaze lands on a target are early triggers, pinches occurring within $350~ms$ after a gaze leaves a target are late triggers). This window was informed by literature on human reflexes \cite{woods2015factors, xbitreactiontest, kosinski2008literature}.

In the second stage, we imported the data into R for statistical modeling. First, we ran a series of repeated measures ANOVA to determine the effect of \textit{Condition} across all metrics. Second, we conducted post-hoc paired-samples t-tests to compare the baseline (\textsc{None}) condition against our heuristic conditions (\textsc{Sticky}, \textsc{Magnetic}, and \textsc{Sticky+ Magnetic}). Additionally, we used within-condition paired t-tests to compare Error Rates to Unaltered Error Rates. This approach provided both hypothesis-driven analysis via t-tests and confirmation of effects via ANOVA.

\begin{figure}
\includegraphics[width=\columnwidth]{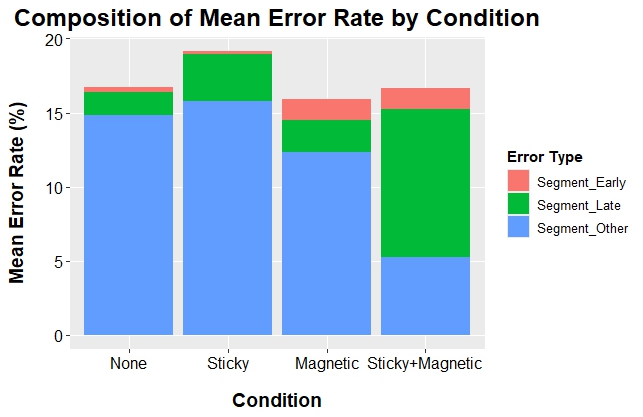}
\caption{\label{fig: errorcomposition} Different types of errors recorded across each study condition (by percentage). Red = Early Trigger errors, Green = Late trigger errors, Blue = All other errors not classified as late or early triggers (i.e., outside the 350 ms window).}
\Description{A stacked bar chart depicting the composition of the average error rates participants recorded throughout the study.

The plot is titled "Composition of Mean Error Rate Across Conditions." It shows four categories on the horizontal axis labeled "Condition", which are None, Sticky, Magnetic, and Sticky+Magnetic, and runs from 0 to 20 on the vertical axis labeled "Mean Error Rate (\%)." In all conditions besides Sticky+Magnetic, the category of Other errors (all errors that were not classified as either late or early triggers) is the most common error type by far, making up around 80 percent of the errors for each condition. In the Sticky+Magnetic condition, the most common error type is Late errors, or late trigger errors, making up about 60 percent of the errors recorded. Magnetic and Sticky+Magnetic are the only conditions that report substantial amounts of Early errors, or early trigger errors, around 12 percent of their total error composition for both. None and Sticky have a marginal slice of errors devoted to Early errors. None, Sticky, and Magnetic each have about 15 percent of their errors devoted to late trigger errors, with Sticky having slightly more than the other two conditions.
}
\end{figure}

\section{Results}

We report the results of statistical tests on nine participants' gaze-and-pinch selections in VR. Due to the small sample size, p-values are reported without correction ($\alpha = .05$) to maintain adequate statistical power while minimizing the risk of false negatives.

We show the full composition of errors across our study in Figure \ref{fig: errorcomposition}. As seen in this figure, absolute error rates varied minimally and non-significantly (see section \ref{sec: Task Performance}) across study conditions. While the difference between \textsc{Sticky} (lowest error rate) and \textsc{Magnetic} (highest error rate) approached significance more closely than other condition pairs, overall impact on the final error count was negligible ($p = 0.34$). Thus, our results investigate specific types of errors and error mitigation differences between the heuristic conditions.

\begin{figure}
\includegraphics[width=\columnwidth]{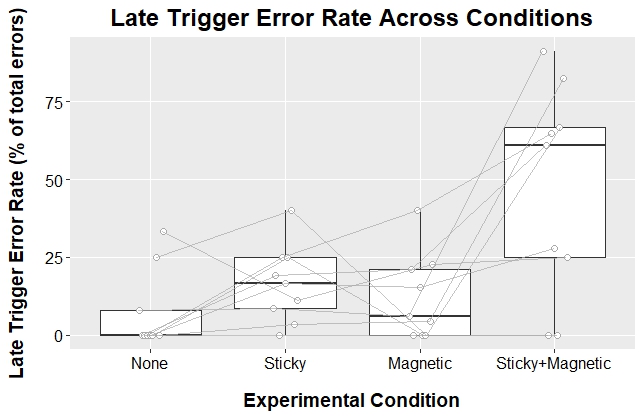}
\caption{\label{fig: latetriggerrate} Late trigger error rates recorded across each study condition (by percentage). Lines connect the points on the plot represented by the same participant, to show the trend of their performance across conditions.}
\Description{A box-and-whisker plot depicting the different late trigger error rates participants recorded throughout the study.

The plot is titled "Late Trigger Error Rate Across Conditions." It shows four categories on the horizontal axis labeled "Conditions", which are None, Sticky, Magnetic, and Sticky+Magnetic, and runs from 0 to 100 on the vertical axis labeled "Late Error Rate." Roughly, None has the smallest spread of points and the lowest error rate, Sticky and Magnetic have similar spreads of points and slightly higher error rates, and Sticky+Magenetic has the largest spread of points and the highest error rate. The median line for the None, Sticky, and Magnetic conditions is about equal to each other (all within 20 percent of each other), while the Sticky+Magnetic median line is by far the highest (around the 62\% mark).
}
\end{figure}

\begin{figure}
\includegraphics[width=\columnwidth]{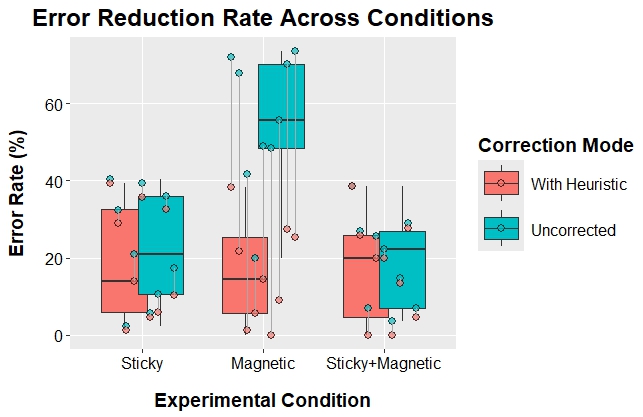}
\caption{\label{fig: errorreduction} Error reduction rates recorded across each study condition (by percentage). Lines connect points represented by the same participant, to show difference between error rates. Solitary red points indicate cases where the error rate was identical with and without heuristics.}
\Description{A box-and-whisker plot depicting the different error reduction rates participants recorded throughout the study.

The plot is titled "Error Reduction Rate Across Conditions." It shows three categories on the horizontal axis labeled "Conditions", which are Sticky, Magnetic, and Sticky+Magnetic, and runs from 0 to 80 on the vertical axis labeled "Error Rate." Across all conditions, the second box plot for "unaltered" or would-be errors is higher than the first box for the final errors recorded with the active heuristic, indicating that the heuristic reduced errors across all conditions. The difference in the Sticky and Sticky+Magnetic is smallest, while the difference between the box plots in the Magnetic condition is vast (with the top of the box plot jumping from roughly 25 percent of final errors to 70 percent would-be errors). The spread of errors across all Conditions is roughly the same, with the spread of Sticky being the widest by a very slight margin. The median lines across all Conditions are similar, aside from the median in the unaltered Magnetic condition, which is high above the rest.
}
\end{figure}

\subsection{Error Reduction} We first analyzed the effectiveness of the heuristics by comparing the final observed error rate against the ``unaltered'' rate (errors that would have occurred without correction). The ANOVA produced a highly significant effect of \textit{Condition} on \textbf{Error Reduction} ($F_{2, 16} = 81.47, p < .001, \eta^2_{G} = .885$), indicating that the heuristics were functionally effective at correcting slips (see Figure \ref{fig: errorreduction}). Post-hoc analysis further confirmed this finding across each study condition. All three conditions showed a significant reduction in errors. The \textsc{Magnetic} condition provided the most substantial improvement, reducing errors by an average of 39.45 mistakes per session ($t(8) = -10.88, p < .001$). Significant reductions were also found for \textsc{Sticky} ($M_{diff} = 3.58, t(8) = -4.69, p = .002$) and \textsc{Sticky+Magnetic} ($M_{diff} = 2.67, t(8) = -3.49, p = .008$).

\subsection{Error Composition (Late vs. Early Triggers)} While overall errors were mitigated, specific error types increased relative to the baseline. We found a significant effect of condition on \textbf{Late Trigger Errors} ($F_{3, 24} = 7.27, p = .001, \eta^2_{G} = .385$), driven primarily by the combined \textsc{Sticky+Magnetic} mode based on our exploratory visuals (see Figure \ref{fig: latetriggerrate}). The ANOVA reported no significant effect on \textbf{Early Trigger Errors}, however ($F_{3, 24} = 1.96, p = .146, \eta^2_{G} = .165$). When running our post-hoc tests, we found that the \textsc{Sticky+ Magnetic} condition resulted in significantly higher late trigger rates ($M = 46.53, SD = 34.24$) than the baseline ($M = 7.37, SD = 12.80, t(8) = -3.26, p = .012$). The \textsc{Sticky} condition showed a non-significant trend toward an increase ($p = .109$). The \textsc{Magnetic} condition also showed a trend toward increased early triggers ($M = 7.72, SD = 9.18$) compared to the baseline ($M = 1.34, SD = 3.32, t(8) = -1.77, p = .114$).

\subsection{Task Performance} \label{sec: Task Performance} The ANOVA did not reach statistical significance for any of our variables related to task performance (\textbf{Throughput}: $F_{3, 24} = 0.47, p = .705, \eta^2_{G} = .011$, \textbf{Selection Time}: $F_{3, 24} = 1.16, p = .346, \eta^2_{G} = .036$, \textbf{Overall Error Rate}: $F_{3, 24} = 0.24, p = .864, \eta^2_{G} = .007$). 

This suggests that the heuristics altered the nature of the errors without negatively impacting user speed or the final accuracy count. Our post-hoc t-tests confirmed this, showing that no significant differences were found between baseline and any assistive condition for Throughput (all $p > .28$), Selection Time (all $p > .19$), or the final Overall Error Rate (all $p > .63$). These results confirm that while the heuristics are powerful tools for error correction, their impact is localized to error composition.

\section{Discussion: User Behavioral Adaptation}

Our results indicate that the introduced heuristics had a profound impact on user behavior, particularly in the \textsc{Magnetic} condition. While absolute error rates remained statistically similar to the baseline—indicating that these heuristics did not fundamentally ``solve" the accuracy issue—the internal behavior of users shifted significantly. We observed that participants' selections during the \textsc{Magnetic} condition was objectively less precise. The amount of would-be errors caught by our tests for error reduction increased dramatically compared to other conditions, showing that users in the \textsc{Magnetic} condition frequently pinched before their gaze reached the target. In this way, they appeared to rapidly adapt to the assistance provided by the Magnetic field, trading specificity when selecting targets for speed.

This explains the high ``Error Reduction'' values observed in the \textsc{Magnetic} condition: users were more likely to pinch further away from the target center—actions that would usually constitute errors—because they correctly anticipated the system would compensate. This suggests that users do not view error correction merely as a safety net, but as an affordance that allows them to offload the cognitive and motor effort of precise targeting. However, this adaption presents a potential drawback, as high reliance on system compensation to handle precision may lead to unintended target selections, particularly in target-dense environments. These adaptations suggest that future correction methods must find a balance—prioritizing the speed users clearly desire while maintaining the agency and precision required for complex tasks. Rather than simply enforcing accuracy by helping users lock onto specific targets or discrete areas within VR, future algorithms should focus on identifying user intent (what the user intends to look at) to facilitate faster selection without sacrificing control.

\section{Conclusion and Future Work}

We present a preliminary investigation into the efficacy of heuristic correction methods—specifically \textsc{Sticky} and \textsc{Magnetic} approaches—for mitigating coordination errors in gaze-and-pinch selections. While the introduction of these heuristics did not result in a statistically significant improvement in throughput or error rates, we revealed critical insights into user behaviors. Namely, the \textsc{Magnetic} condition encouraged users to adopt a faster, less precise selection strategy, ``offloading'' the burden of accuracy onto the system.

This behavioral shift—sacrificing precision for speed when assistance is present—suggests that the future of gaze-and-pinch lies not in enforcing precision, but in predicting user intent. It also suggests a potentially significant trade-off: while heuristics may mitigate coordination errors, they may also encourage a reliance on the system that masks underlying performance bottlenecks. Future work should employ subjective evaluations, gathering users' perceived speed and convenience with the system via interviews to interpret where significant bottlenecks are felt. Explicit measures, such as the NASA-TLX, should also be used to quantify whether user ``offloading" reduces cognitive workload or introduces new frustrations. Finally, exploring the parameter spaces of these heuristics, such as varying the timing window of \textsc{Sticky} effects and sizes of the \textsc{Magnetic} field, will be essential to optimizing user interactions via these heuristics. Ultimately, this work highlights that future interfaces must look beyond objective throughput alone, and should aim to explore and accommodate users' behavioral changes to make VR interactions robust and fluid.

\bibliographystyle{ACM-Reference-Format}
\bibliography{references}

\end{document}